\documentclass[a4paper,12pt]{article}
\usepackage{a4wide}

\usepackage[latin1]{inputenc}
\usepackage{dsfont}
\usepackage{amsfonts}
\usepackage{amsmath}
\usepackage{amssymb}
\usepackage{amsthm}
\usepackage{graphicx}
\usepackage{booktabs}
\usepackage{color}
\usepackage{ulem}

\newcommand{\be}{\begin{equation}}
\newcommand{\ee}{\end{equation}}
\newcommand{\beq}{\begin{equation}}
\newcommand{\eeq}{\end{equation}}
\newcommand{\bea}{\begin{eqnarray}}
\newcommand{\eea}{\end{eqnarray}}

\begin{document}

\begin{titlepage}

\vspace*{-15mm}
\vspace*{0.7cm}

\begin{center}
{\large {\bf Large neutrino mixing angle $\theta^{\text{MNS}}_{13}$ and quark-lepton mass ratios \\[1mm]  in unified flavor models}}\\[8mm]

Stefan Antusch$^{\star\dagger}$\footnote{Email: \texttt{stefan.antusch@unibas.ch}},~
Vinzenz Maurer$^{\star}$\footnote{Email: \texttt{vinzenz.maurer@unibas.ch}}
\end{center}

\vspace*{0.20cm}

\centerline{$^{\star}$ \it
Department of Physics, University of Basel,}
\centerline{\it
Klingelbergstr.\ 82, CH-4056 Basel, Switzerland}

\vspace*{0.4cm}

\centerline{$^{\dagger}$ \it
Max-Planck-Institut f\"ur Physik (Werner-Heisenberg-Institut),}
\centerline{\it
F\"ohringer Ring 6, D-80805 M\"unchen, Germany}

\vspace*{1.2cm}

\begin{abstract}
We analyze how a large value of the leptonic mixing angle $\theta^{\text{MNS}}_{13}$ can be generated via charged lepton corrections in unified flavor models, using novel combinations of Clebsch-Gordan factors for obtaining viable quark-lepton mass ratios for the first two families. We discuss how these Clebsch-Gordan factors affect the relations between down-type quark mixing and charged lepton mixing in SU(5) grand unified theories and Pati-Salam unified models and calculate the resulting possible predictions for $\theta_{13}^{\text{MNS}}$ for models with $\theta_{13}^\nu, \theta_{13}^e \ll \theta_{12}^e$. While symmetric mass matrices with zero (1,1)-elements always lead to comparatively small $\theta^{\text{MNS}}_{13}\approx  2.8^\circ$, we find novel combinations of Clebsch-Gordan factors for nonsymmetric mass matrices which can yield $\theta^{\text{MNS}}_{13}\approx  5.1^\circ$, $6.1^\circ$, $7.6^\circ$ or $10.1^\circ $, as favored by the current experimental hints for large $\theta^{\text{MNS}}_{13}$. We discuss applications to classes of models with underlying tribimaximal or bimaximal mixing in the neutrino sector.  
\end{abstract}


\end{titlepage}

\newpage

\section{Introduction}

The discovery of large, close to tribimaximal lepton mixing has lead to a revolution in flavor model building. Over the last years, a variety of models has been constructed \cite{FlavourRev} which feature tribimaximal mixing \cite{HPS} in the neutrino mass matrix.  Furthermore, alternatives for underlying mixing patterns are discussed as well, for instance bimaximal neutrino mixing (for model realizations, see, e.g.\ \cite{Altarelli:2009gn}). Many of them, regardless of the specific seesaw type, have in common that the 1-3 mixing $\theta_{13}^{\nu}$ in the neutrino sector (almost) vanishes. In the context of grand unified theories (GUTs) and quark-lepton unified models, such a special mixing pattern in the neutrino sector typically receives, however, corrections from charged lepton mixing which is often dominated by a sizable $\theta_{12}^e$. With $\theta_{12}^e \gg \theta_{13}^e, \theta_{13}^\nu$, a leptonic (MNS) mixing angle $\theta_{13}$ of the size
\be
\theta_{13}^{\text{MNS}}  \approx \frac{1}{\sqrt{2}} \theta_{12}^e
\ee
is generated (see, e.g.\ \cite{CLcorrections}).\footnote{We remark that the relation holds exactly only for $\theta_{23}=45^\circ$ and we use this form here for simplicity. The more precise relation is given by $\theta_{13}^{\text{MNS}}  = \theta_{12}^e \sin(\theta_{23}^{\text{MNS}})$.}  
The charged lepton mixing $\theta_{12}^e$, in turn, is often related to $\theta_{12}^d$ via GUT relations. The down quark mixing $\theta_{12}^d$ is in principle model-dependent, however, since the up-type quark masses are more hierarchical than the down-type quark ones, in many classes of models the down quark mixing is larger than the up-quark mixing and the relation $\theta_{12}^d \approx \theta_C \approx 13^\circ$ holds to some approximation. 
A well-known case, realized in many models in the literature, appears when the quark-lepton mass ratio $m_\mu/m_s = 3$ is realized via a ``Georgi-Jarlskog'' \cite{GJ} Clebsch factor of $3$  in the (2,2)- element of $Y_e$. 
Then, $\theta_{12}^e = \theta_{12}^d / 3 \approx  \theta_C / 3$ holds and the ubiquitous prediction
\be
\theta_{13}^{\text{MNS}}  = \frac{1}{3\sqrt{2}} \theta_C \approx 3^\circ
\ee
arises. 

However, if the current experimental hints at a large value $\theta_{13}^{\text{MNS}} $, e.g.\ from T2K \cite{:2011sj} (for a recent global fit, see \cite{Fogli:2011qn}), get confirmed, this prediction would be disfavored. In this article we will therefore discuss how alternative Clebsch factors and GUT scale quark-lepton mass relations, including the novel possibilities proposed recently in \cite{Antusch:2009gu}, can influence the predictions for $\theta_{13}^{\text{MNS}} $ in unified flavor models.

\section{The relation between $\theta_{12}^e$ and $\theta_{12}^d$ in unified models}\label{sec:t12et12d}
In the following, we will assume that $\theta_{12}^d \approx \theta_C \approx 13^\circ$ \cite{Amsler:2008zzb}, as discussed above, and that $Y_d$ and $Y_u$ are hierarchical. To be explicit, one may simply take the case that $Y_u$ is (approximately) diagonal. 
We will below include formulae for general $\theta_{12}^d$ such that the results can readily be adopted to other cases as well.  Furthermore, we will focus on classes of models where the mass matrices feature zero (or negligible) (1,1)-elements. In the following discussion we will distinguish between the case of unified models with Pati-Salam (PS) gauge group $G_{422}=$SU(4)$_C\times$SU(2)$_L\times$SU(2)$_R$ and Grand Unified models based on SU(5).

\subsection{Pati-Salam unified models}\label{sec:PS}
In PS unified models, under the above-specified assumptions, $Y_e$ is related to $Y_d$, with each element depending on a Clebsch factor $c_i$.\footnote{We assume general, not necessarily symmetric, Yukawa matrices and that one single operator dominates each matrix element.} Focusing on the 2$\times$2 submatrix of the first two families, we can write
\be
Y_d = \begin{pmatrix} 
0&b\\
a&c
\end{pmatrix}\:
\Rightarrow \quad
Y_e = \begin{pmatrix} 
0& c_b \, b\\
 c_a\, a&c_c \, c
\end{pmatrix}.
\ee 
Note that we are using the same symbols for the full Yukawa matrices and the 2$\times$2 submatrices here and in the following. 
From the above matrices we can immediately read off the approximate relation
\be
\theta_{12}^e \approx \left\vert \frac{c_b}{c_c} \right\vert \theta_{12}^d \; , 
\ee
and thus 
\be
\theta_{13}^{\text{MNS}} \approx \frac{1}{\sqrt{2}} \left\vert\frac{c_b}{c_c}  \right\vert \theta_{12}^d  \approx 9.2^\circ  \left\vert\frac{c_b}{c_c} \right\vert \frac{\theta_{12}^d}{\theta_C}\;.
\ee

\subsection{SU(5) GUTs}\label{sec:SU(5)}
In SU(5) GUTs the situation is somewhat more tricky, since $Y_e$ is related not to $Y_d$ but to $Y_d^T$. Without loss of generality we can now write
\be
Y_d = \begin{pmatrix} 
0&b\\
a&c
\end{pmatrix}\:
\Rightarrow \quad
Y_e = 
\begin{pmatrix} 
0& c_b \, b\\
 c_a\, a&c_c \, c
\end{pmatrix}^T
=
\begin{pmatrix} 
0& c_a \, a\\
 c_b\, b&c_c \, c
\end{pmatrix}
\ee 
which results in
\be
\theta_{12}^e \approx \left\vert\frac{c_a \, a}{c_c \, c}  \right\vert\;.
\ee
To continue from here we can use
\be
m_\mu \approx \left\vert c_c \, c  \right\vert \;,\quad m_e \approx \left\vert\frac{c_a \, a\: c_b \,b}{c_c \, c}  \right\vert
\ee
to obtain
\be
\theta_{12}^e \approx \frac{m_e}{m_\mu} \left\vert \frac{c_c}{c_b}  \right\vert \frac{1}{\theta_{12}^d}  \;,
\ee
and thus 
\be
\theta_{13}^{\text{MNS}}  \approx \frac{1}{\sqrt{2}} \frac{m_e}{m_\mu}  \left\vert \frac{c_c}{c_b}  \right\vert \frac{1}{\theta_{12}^d}  \approx 
0.84^\circ \left\vert \frac{c_c}{c_b}  \right\vert \frac{\theta_C}{\theta_{12}^d}\;.
\ee
We see that, compared to the Pati-Salam case, we have the inverse dependence on the Clebsch factors as well as on $\theta_{12}^d$.\footnote{We note that this also implies, for cases where $\theta_{12}^d \approx \theta_C$ may not hold anymore, that $\theta_{12}^d < \theta_C$ provides another way of obtaining larger $\theta_{13}^{\text{MNS}}$ in SU(5), whereas in Pati-Salam models we would need $\theta_{12}^d > \theta_C$ for increasing $\theta_{13}^{\text{MNS}}$.}

\section{Quark-lepton mass relations and possible Clebsches}\label{sec:Clebsches}
The quark-lepton mass ratios at the GUT scale, again under the assumptions specified in the previous section, have the following approximate expressions in terms of the Clebsch factors:
\be
\frac{m_\mu}{m_s} \approx \left\vert c_c  \right\vert \;,\quad \frac{m_e}{m_d} \approx \left\vert\frac{c_a c_b}{c_c}  \right\vert \;.
\ee
Note that when we refer to the mass ratios at the GUT scale, we take them as equivalent to the ratios of the running Yukawa couplings, i.e., we always use $m_i := y_i v$ with the Higgs vacuum expectation value $v$ fixed by its value at the electroweak scale. 
Using the results from \cite{Antusch:2009gu} we will now consider the favorable Clebsch factors\footnote{In principle, also $|c_c| = 2$ in PS and $|c_c| = 9$ in PS and SU(5) may work as well for some classes of supersymmetric (SUSY) models. However, they would require large SUSY threshold corrections (above 40\%) to be consistent with data. It is straightforward to generalize our analysis also to these cases.} 
\be
\left\vert c_c  \right\vert \in \{3,\tfrac{9}{2},6\} \:,
\ee 
where the first one is the GJ factor \cite{GJ} and where the latter two have been proposed in \cite{Antusch:2009gu}. We note that while the Clebsch factor $3$ is available in PS as well as in SU(5), $9/2$ and $6$ are only available in SU(5)-like setups. 
The viability of these Clebsch factors depends on the SUSY breaking scenario (more generally on the SUSY model parameters).

Assuming SUSY threshold corrections such that the Clebsches $c_c$ yield a ratio $\frac{m_\mu}{m_s}$ within the allowed experimental range, we can now calculate the allowed region for the product $| c_a c_b |$ such that $\frac{m_e}{m_d}$ is in the (1$\sigma$) allowed region as well. For this, we make the simplifying assumption that the SUSY threshold corrections are almost the same for the first two generations, which is usually a good approximation when there are universal boundary conditions for the soft SUSY breaking parameters at the GUTs scale. For the Clebsch factors $c_c$ under consideration we obtain:
\bea
\left\vert c_c  \right\vert &=& 3 \; \Rightarrow \quad | c_a c_b |\in [ 0.4,  1.8] \;,\\
\left\vert c_c  \right\vert &=& \tfrac{9}{2} \; \Rightarrow \quad |c_a c_b| \in [ 0.9, 4.1] \;,\\
\left\vert c_c  \right\vert &=& 6 \; \Rightarrow \quad |c_a c_b| \in [ 1.6, 7.2 ] \;.
\eea
If we now take for the PS case, see e.g.\ \cite{Allanach:1996hz}, the possible Clebsches (to be applied to $|c_c| =3 $)
\be
\left\vert c_a  \right\vert , \left\vert c_b  \right\vert \in \{\tfrac{3}{4},1,2,3,9\} \:,
\ee 
and for the SU(5) cases, see e.g.\ \cite{Antusch:2009gu}, (to be applied to $|c_c| \in \{3,\tfrac{9}{2},6\} $)
\be
\left\vert c_a \right\vert , \left\vert c_b \right\vert  \in \{\tfrac{1}{2},1,\tfrac{3}{2},3,\tfrac{9}{2},6,9\} \:,
\ee
we obtain that $\vert\tfrac{c_c}{c_b} \vert $ can take the (discrete) values
\bea
\left\vert c_c  \right\vert &=& 3\; \Rightarrow  \quad
\begin{cases}
\mbox{In PS:} \;\: & \vert\tfrac{c_b}{c_c}  \vert \in \{ \tfrac{1}{4} , \tfrac{1}{3} , \tfrac{2}{3} \} \;, \\
\mbox{In SU(5):}\;\:&  \vert\tfrac{c_c}{c_b} \vert  \in \{ 1,2,3,6\} \;, 
\end{cases}\\
\left\vert c_c  \right\vert &=& \tfrac{9}{2}\; \Rightarrow \quad \vert\tfrac{c_c}{c_b}\vert \in \{  \tfrac{3}{4}  , 1,  \tfrac{3}{2}, 3 ,  \tfrac{9}{2} ,9 \} \;, \\
\left\vert c_c \right\vert  &=& 6 \; \Rightarrow \quad \vert\tfrac{c_c}{c_b}\vert \in \{  \tfrac{2}{3} , 1,  \tfrac{4}{3} , 2 , 4 , 6 , 12 \} \;.
\eea

\section{Predictions for $\theta_{13}$}\label{sec:predictions}

When we assume that $\theta_{12}^d \approx \theta_C $ we can translate the above-derived ratios of $\tfrac{c_b}{c_c} $ and $\tfrac{c_c}{c_b}$ into possible GUT predictions for $\theta_{13}^{\text{MNS}} $:
 \bea
\left\vert c_c  \right\vert &=& 3\; \Rightarrow 
\begin{cases}
\mbox{In PS:} \;\: \quad \; \theta_{13}^{\text{MNS}}  \in \{ 2.3^\circ, 3.1^\circ , 6.1^\circ \} \;, \\
\mbox{In SU(5):}\;\: \theta_{13}^{\text{MNS}}  \in \{ 0.8^\circ, 1.7^\circ, 2.5^\circ, 5.1^\circ\} \;, 
\end{cases}\\
\left\vert c_c  \right\vert &=& \tfrac{9}{2}\; \Rightarrow \theta_{13}^{\text{MNS}}  \in \{  0.6^\circ , 0.8^\circ,  1.3^\circ, 2.5^\circ , 3.8^\circ ,7.6^\circ \} \;, \\
\left\vert c_c  \right\vert &=& 6 \; \Rightarrow \theta_{13}^{\text{MNS}}  \in \{  0.6^\circ , 0.8^\circ, 1.1^\circ , 1.7^\circ , 3.4^\circ, 5.1^\circ , 10.1^\circ \} \;.
\eea
We can see that there are several cases which feature a comparatively large $\theta_{13}^{\text{MNS}} $. The corresponding combinations of Clebsch factors are given in Table \ref{tab:t13}. Unified flavor models which use one of these combinations of Clebsches can thus realize a large value of $\theta_{13}^{\text{MNS}}$ despite zero (small) $\theta_{13}^\nu$.

\begin{table}
\begin{center}
\begin{tabular}{|l|c|p{4.3mm}p{4.3mm}p{4.3mm}|p{4.3mm}p{4.3mm}p{4.3mm}|ccc|ccc|}
\hline
\vphantom{$\dfrac{v}{v}$}$\vert c_c \vert$   & $3$ in PS     &
\multicolumn{3}{|c|}{$3$ in SU(5)}                  &
\multicolumn{3}{|c|}{$\tfrac{9}{2}$ in SU(5)}                &
\multicolumn{6}{|c|}{$6$ in SU(5)} \\
\hline
\vphantom{$\dfrac{v}{v}$}$\theta_{13}$       & $6.1^\circ$   &
\multicolumn{3}{|c|}{$5.1^\circ$}                   &
\multicolumn{3}{|c|}{$7.6^\circ$}                   &
\multicolumn{3}{|c|}{$5.1^\circ$} & \multicolumn{3}{|c|}{$10.1^\circ$} \\
\hline
\vphantom{$\dfrac{v}{v}$}$\vert c_a \vert$ & $\tfrac{3}{4}$& $1$ & $\tfrac{3}{2}$ & $3$ & $3$ & $\tfrac{9}{2}$ & $6$ & $3$ & $\tfrac{9}{2}$ & $6$ & $\tfrac{9}{2}$ & $6$ & $9$ \\
\vphantom{$\dfrac{v}{v}$}$\vert c_b \vert$   & $2$ & $\tfrac{1}{2}$ & $\tfrac{1}{2}$ & $\tfrac{1}{2}$ & $\tfrac{1}{2}$ & $\tfrac{1}{2}$ & $\tfrac{1}{2}$ & $1$ & $1$ & $1$ & $\tfrac{1}{2}$ & $\tfrac{1}{2}$ & $\tfrac{1}{2}$ \\
\hline
\end{tabular}
\end{center}
\caption{\label{tab:t13} Examples for prediction for large $\theta_{13}^{\text{MNS}}$ which arise with Clebsch factors $c_c$ and $c_a,c_b$ and under the conditions specified in the main text.}
\end{table}

We have assumed here that charged lepton mixing is the dominant source of $\theta_{13}^{\text{MNS}}$. We note that, of course, in explicit models various additional corrections to $\theta_{13}^{\text{MNS}}$ may have to be taken into account, e.g.\ from RG running (see e.g.\ \cite{Antusch:2005gp}), from canonical normalization corrections (c.f.\ \cite{Antusch:2007ib}), or from the next order in small angle expansion for the charged lepton mixing effects. For more accurate predictions, a careful model analysis is required.

\section{Special case: Symmetric mass matrices} \label{sec:symmetric}
A special case is given by the possibility of symmetric mass matrices $Y_d$ and $Y_e$, where 
\be
Y_e = 
\begin{pmatrix} 
0& c_a \, a\\
 c_a\, a&c_c \, c
\end{pmatrix}.
\ee 
In this case $\theta^e_{12}$ is fully determined by $m_e$ and $m_\mu$:
\be
\theta^e_{12} = \sqrt{\frac{m_e}{m_\mu}}\;.
\ee 
This leads to the prediction 
\be
\theta_{13}^{\text{MNS}}  = \frac{1}{\sqrt{2}}\sqrt{\frac{m_e}{m_\mu}} \approx 2.8^\circ\;.
\ee
We emphasize that this prediction is independent of Clebsch factors and independent of GUT relations. It is featured by many existing models in the literature. In fact, in unified models and under the conditions we have specified above, the GUT scale values of $\theta^d_{12}$ and of $m_s,m_d$ are now predicted by the Clebsch factors and the charged lepton masses (since the charged lepton masses are known most precisely and dominate the fit to the data). With $c_c$ and $c_a$ as defined above, one obtains for instance: 
\be
\theta^d_{12} \approx  \sqrt{\frac{m_e}{m_\mu}} \left\vert\frac{c_c}{c_a}  \right\vert  \approx 3.9^\circ  \left\vert\frac{c_c}{c_a}  \right\vert \;.
\ee
On the other hand, as in this case a comparatively small $\theta_{13}^{\text{MNS}} $ is predicted, this scenario would be disfavored if the recent experimental hints \cite{:2011sj,Fogli:2011qn} towards a rather sizable reactor mixing angle get confirmed.

\section{Special neutrino mixing patterns, large $\theta_{13}^{\text{MNS}}$ and the lepton mixing sum rule}
Finally, we would like to comment on the effect of $\theta_{12}^e$-dominated charged lepton corrections on the solar mixing angle $\theta_{12}$, with a special focus on possibly large $\theta_{13}^{\text{MNS}} $. In fact, under the assumptions specified above, the lepton mixing sum rule \cite{CLcorrections,sumrule}
\be
\theta_{12}^{\text{MNS}}  - \theta_{13}^{\text{MNS}}  \cos (\delta_{\text{MNS}} ) = \theta_{12}^\nu\;,
\ee  
with $\theta_{12}^\nu$ being the prediction for the 1-2 mixing in the neutrino sector, holds. This means, when we assume underlying tribimaximal neutrino mixing with $\sin(\theta_{12}^\nu)=1/\sqrt{3}$ then large $\theta_{13}^{\text{MNS}} $ can only be consistent for (approximately) $\delta_{\text{MNS}}  \in \{90^\circ , 270^\circ\}$.\footnote{Such specific predictions for $\delta_{\text{MNS}}$ may emerge from flavor models with $\mathbb{Z}_2$ or $\mathbb{Z}_4$ shaping symmetries to explain a right-angled CKM unitarity triangle (with $\alpha \approx 90^\circ$), as discussed recently in \cite{Antusch:2011sx}.}  

On the other hand, for bimaximal neutrino mixing with $\theta_{12}^\nu = 45^\circ$ the experimental data even requires large $\theta_{13}^{\text{MNS}} $ close to the present bounds, together with $\delta_{\text{MNS}}  \approx 180^\circ$. The leptonic mixing sum rule would then simply turn into $\theta^{\text{MNS}}_{12} + \theta^{\text{MNS}}_{13}  = 45^\circ$. With $\theta^{\text{MNS}}_{13} \approx 10.1^\circ$, obtainable in SU(5) scenarios with $|c_c| = 6$ and $(|c_a| , |c_b| ) = (  \tfrac{9}{2}, \tfrac{1}{2})$, $(6, \tfrac{1}{2})$ and $ ( 9, \tfrac{1}{2})$, this would be in excellent agreement with the current experimental data.

In general, using the sum rule, a precise measurement of $\theta_{12}^{\text{MNS}}, \theta_{13}^{\text{MNS}} $ and $\delta_{\text{MNS}} $ may provide a smoking gun signal for a specific neutrino mixing pattern.

\section{Summary and conclusions} 
In this article, we have discussed how a large value of the leptonic mixing angle $\theta^{\text{MNS}}_{13}$ can be generated via charged lepton corrections in unified models, using novel combinations of Clebsch factors which can lead to viable quark-lepton mass ratios for the first two families. 
Our approach may thus provide a way to build unified flavor models consistent with current experimental data (which hints at large  $\theta^{\text{MNS}}_{13}$) and with an underlying special mixing pattern in the neutrino sector (such as tribimaximal or bimaximal mixing). 

Predictions for $\theta^{\text{MNS}}_{13}$ can arise as follows:
If the conditions $\theta_{13}^\nu,\theta_{13}^e\ll \theta_{12}^e$ are satisfied, the leptonic mixing angle $\theta^{\text{MNS}}_{13}$ is given by
\be
\theta_{13}^{\text{MNS}}  \approx \frac{1}{\sqrt{2}} \theta_{12}^e \;,
\ee
i.e., via the charged lepton 1-2 mixing $\theta_{12}^e$. 
In unified flavor models, the down-type quark Yukawa matrix $Y_d$ and the charged lepton Yukawa matrix $Y_e$ are typically generated by the same operators, since quarks and leptons are in joint representations of the unifying gauge group. This implies that their elements may differ only by group theoretical Clebsch factors from the breaking of the extended gauge symmetry. For instance, in classes of models where the Yukawa matrices have a zero texture in their (1,1)-elements and where $\theta_{12}^d \approx \theta_C$, it follows that the value of $\theta_{12}^e$ satisfies (with the Clebsch factors $c_c$ and $c_b$):
\be
\theta_{12}^e \approx  
\begin{cases}
\vert \frac{c_b}{c_c} \vert \theta_C & \mbox{(in PS)}\;,\\
\frac{m_e}{m_\mu} \vert \frac{c_c}{c_b} \vert \frac{1}{\theta_C}&\mbox{(in SU(5))\;.}
\end{cases}
\ee 
To use these formulae for calculating the predictions for $\theta^{\text{MNS}}_{13}$, we have determined possible consistent combinations of Clebsch factors in SU(5) and Pati-Salam unified theories which lead to viable mass ratios of the first two families of down-type quarks and charged leptons. In addition to the well-known Clebsch factor $|c_c |= 3$, corresponding to the ubiquitous relation $m_\mu/m_s = 3$ \cite{GJ}, we also included the recently discussed novel possibilities $|c_c| = 9/2$ and $|c_c|=6$, which lead to $m_\mu/m_s = \tfrac{9}{2}$ or $6$ \cite{Antusch:2009gu}, respectively, in our discussion.

We then calculated the corresponding predictions for $\theta^{\text{MNS}}_{13}$. We found that certain combinations of Clebsch factors can lead to predictions $\theta^{\text{MNS}}_{13} \in \{5.1^\circ,6.1^\circ,7.6^\circ,10.1^\circ \} $ (c.f. Table \ref{tab:t13}), which can accommodate recent experimental hints for large $\theta^{\text{MNS}}_{13}$. Charged lepton corrections to $\theta_{12}^\nu$ are suppressed for $\delta_{\text{MNS}} \in \{90^\circ,270^\circ\}$ and maximal for $\delta_{\text{MNS}} \in \{0^\circ,180^\circ\}$, according to the lepton mixing sum rule $\theta^{\text{MNS}}_{12} - \theta^{\text{MNS}}_{13} \cos (\delta_{\text{MNS}}) = \theta_{12}^\nu$ \cite{CLcorrections,sumrule}. In models capable of predicting maximal CP violation (c.f.\ \cite{Antusch:2011sx}), large $\theta_{13}$ may thus be realized while preserving the favorable prediction $\sin(\theta^{\text{MNS}}_{12}) \approx \sin(\theta^\nu_{12}) \approx 1/\sqrt{3}$ from a possible underlying tribimaximal neutrino mixing pattern. On the other hand, for models with $\delta_{\text{MNS}} =180^\circ $, bimaximal neutrino mixing may be realized and the leptonic mixing sum rule would then simply turn into $\theta^{\text{MNS}}_{12} + \theta^{\text{MNS}}_{13}  = 45^\circ$. With $\theta^{\text{MNS}}_{13} \approx 10.1^\circ$, obtainable in SU(5) scenarios with $|c_c| = 6$ and $(|c_a| , |c_b| ) = (  \tfrac{9}{2}, \tfrac{1}{2})$, $(6, \tfrac{1}{2})$ and $ ( 9, \tfrac{1}{2})$, this would be in excellent agreement with the current experimental data.

In summary, we have discussed possibilities for realizing a large value of $\theta^{\text{MNS}}_{13}$ in unified flavor models using novel combinations of Clebsch factors to obtain consistent quark-lepton mass ratios for the first two families of down-type quarks and charged leptons. While $\theta^{\text{MNS}}_{13}\approx 5^\circ$ or $6^\circ$ can be realized in SU(5) and in Pati-Salam models for $|c_c| = 3$, as well as in SU(5) for $|c_c| = 6$, the largest values $\theta^{\text{MNS}}_{13}\approx 7.6^\circ$ and $10.1^\circ$ were obtained with $|c_c| = \tfrac{9}{2}$ and $6$, respectively, and in both cases with $|c_b| = \tfrac{1}{2}$.\footnote{For ways to obtain these Clebsches from effective GUT operators, realized by messenger fields, see \cite{Antusch:2009gu}.} If the experimental hints for such large $\theta^{\text{MNS}}_{13}$ should get confirmed, these Clebsch factors could become favorable building blocks for future unified flavor models.

\section*{Acknowledgments}
This project was supported by the Swiss National Science Foundation.
S.~A.\ also acknowledges support by the DFG cluster of excellence ``Origin and Structure of the Universe''.


\begin{thebibliography}{99}

\bibitem{FlavourRev}
 For reviews with extensive lists of references, see e.g.:
 G.~Altarelli and F.~Feruglio,
 Rev.\ Mod.\ Phys.\  {\bf 82} (2010) 2701
 [arXiv:1002.0211];
 H.~Ishimori, T.~Kobayashi, H.~Ohki, Y.~Shimizu, H.~Okada and M.~Tanimoto,
 Prog.\ Theor.\ Phys.\ Suppl.\  {\bf 183} (2010) 1
 [arXiv:1003.3552].
G.~Altarelli,
arXiv:1011.5342.

\bibitem{HPS}
P.~F.~Harrison, D.~H.~Perkins and W.~G.~Scott,
Phys.\ Lett.\ B {\bf 530} (2002) 167 [arXiv:hep-ph/0202074].

\bibitem{Altarelli:2009gn}
  G.~Altarelli, F.~Feruglio and L.~Merlo,
  JHEP {\bf 0905} (2009) 020
  [arXiv:0903.1940 [hep-ph]];
  D.~Meloni,
  JHEP {\bf 1110} (2011) 010
  [arXiv:1107.0221 [hep-ph]].


\bibitem{CLcorrections}
S.~Antusch and S.~F.~King,
Phys.\ Lett.\ B {\bf 631} (2005) 42
[arXiv:hep-ph/0508044];
 
 
 \bibitem{GJ}
  H.~Georgi and C.~Jarlskog,
  Phys.\ Lett.\  B {\bf 86} (1979) 297.

\bibitem{:2011sj}
   T2K Collaboration,
  Phys.\ Rev.\ Lett.\  {\bf 107} (2011) 041801
  [arXiv:1106.2822 [hep-ex]].

\bibitem{Fogli:2011qn}
  G.~L.~Fogli, E.~Lisi, A.~Marrone, A.~Palazzo and A.~M.~Rotunno,
  Phys.\ Rev.\ D {\bf 84} (2011) 053007
  [arXiv:1106.6028 [hep-ph]].

 
\bibitem{Antusch:2009gu}
S.~Antusch and M.~Spinrath,
Phys.\ Rev.\  D {\bf 79} (2009) 095004
[arXiv:0902.4644 [hep-ph]].
  
\bibitem{Amsler:2008zzb}
C.~Amsler {\it et al.}  [Particle Data Group],
Phys.\ Lett.\  B {\bf 667} (2008) 1.


\bibitem{Allanach:1996hz}
  B.~C.~Allanach, S.~F.~King, G.~K.~Leontaris, S.~Lola,
  Phys.\ Rev.\  {\bf D56 } (1997)  2632-2655.
  [hep-ph/9610517]. 

\bibitem{Antusch:2005gp}
S.~Antusch, J.~Kersten, M.~Lindner, M.~Ratz and M.~A.~Schmidt,
JHEP {\bf 0503} (2005) 024
[arXiv:hep-ph/0501272].


\bibitem{Antusch:2007ib}
S.~Antusch, S.~F.~King and M.~Malinsky,
Phys.\ Lett.\  B {\bf 671} (2009) 263
[arXiv:0711.4727 [hep-ph]];
 S.~Antusch, S.~F.~King, M.~Malinsky,
 JHEP {\bf 0805 } (2008)  066.
 [arXiv:0712.3759 [hep-ph]];
S.~Antusch, S.~F.~King and M.~Malinsky,
Nucl.\ Phys.\  B {\bf 820} (2009) 32
[arXiv:0810.3863 [hep-ph]].

\bibitem{sumrule}
S.~F.~King,
JHEP {\bf 0508} (2005) 105
[arXiv:hep-ph/0506297];
I.~Masina,
Phys.\ Lett.\  B {\bf 633} (2006) 134
[arXiv:hep-ph/0508031];
S.~Antusch, P.~Huber, S.~F.~King and T.~Schwetz,
JHEP {\bf 0704} (2007) 060
[arXiv:hep-ph/0702286].

\bibitem{Antusch:2011sx}
  S.~Antusch, S.~F.~King, C.~Luhn and M.~Spinrath,
  Nucl.\ Phys.\  B {\bf 850} (2011) 477
  [arXiv:1103.5930 [hep-ph]].




\end{thebibliography}
\end{document}